\documentclass[
showpacs,preprintnumbers,amsmath,amssymb]{revtex4}
\usepackage{amsmath}
\usepackage{graphicx}
\usepackage{subfig}
\usepackage{hyperref}
\usepackage{amssymb}
\usepackage[english]{babel}
\usepackage{epsfig}
\usepackage{wasysym}
\usepackage{graphicx}
\usepackage{indentfirst}
\usepackage{amsmath}
\usepackage{amsfonts}
\usepackage{amssymb}
\usepackage{enumerate}

\hypersetup{colorlinks,citecolor=green,filecolor=magenta,linkcolor=red,urlcolor=cyan,pdftex}

\newcommand{\be}{\begin{equation}}
\newcommand{\ee}{\end{equation}}
\newcommand{\bea}{\begin{eqnarray}}
\newcommand{\eea}{\end{eqnarray}}
\newcommand{\beaa}{\begin{eqnarray*}}
\newcommand{\eeaa}{\end{eqnarray*}}

\begin{document}


\title{Cosmological Evolution in $f(R,T)$ theory with Collisional Matterbnbnb}

 \author{E. H. Baffou$^{(a)}$\footnote{e-mail:baffouh.etienne@yahoo.fr},  M. J. S. Houndjo$^{(a,b)}$\footnote{e-mail:
  sthoundjo@yahoo.fr}, M. E. Rodrigues$^{(d,e)}$\footnote{e-mail: esialg@gmail.com}, A. V. Kpadonou$^{(c)}$\footnote{e-mail: vkpadonou@gmail.com} and J. Tossa$^{(a)}$\footnote{e-mail: joel.tossa@imsp-uac.org} }

  
\affiliation{$^a$ \, Institut de Math\'{e}matiques et de Sciences Physiques (IMSP), 01 BP 613,  Porto-Novo, B\'{e}nin\\
$^{b}$\, Facult\'e des Sciences et Techniques de Natitingou - Universit\'e de Parakou - B\'enin \\
$^{c}$\, Ecole Normale Sup\'erieure de Natitingou - Universit\'e de Parakou - B\'enin\\ 
$^{d}$ \, Faculdade de F\'isica, PPGF, Universidade Federal do Par\'a, 66075-110, Bel\'em, Par\'a, Brazil.\\   
$^{e}$\,Faculdade de Ci\^encias Exatas e Tecnologia, Universidade Federal do Par\'a - Campus Universit\'ario 
de Abaetetuba, CEP 68440-000, Abaetetuba, Par\'a, Brazil }  

\begin{abstract}
We study the evolution of the cosmological parameters, namely, the deceleration parameter $q(z)$ and the parameter of effective equation of state in a universe containing, besides the ordinary matter and dark energy, a self-interacting (collisional matter), in the generalized $f(R,T)$ theory of gravity, where $R$ and $T$ are the curvature scalar and the trace of the energy-momentum tensor, respectively.  We use the generalized FRW equations, the equation of continuity and obtain a differential equation in $H(z)$, and solve it numerically for studying the evolution of the cosmological parameters. Two $f(R,T)$ models are considered and the results with collisional matter are compared with the ones of the $\Lambda$CDM model, and also with the model where there exists only non-collisional matter. The curves show that the models are acceptable because the values found for $w_{eff}$ are consistent with the observational data. 
\end{abstract}


\maketitle 

\section{Introduction}

It is well known nowadays that our current universe is experiencing an accelerated expansion \cite{eti14}-\cite{eti20}. There are two general ways to explain this accelerated expansion of the universe \cite{baf1}-\cite{baf5}. 
The first way is considering that the universe is essentially filled by an exotic fluid with negative pressure, responsible of it acceleration, called the dark energy. The second way is modifying the gravitational action from the General Relativity (GR) without the need of the dark energy, firstly considered by 1970 \cite{baf6}. Buchdahl has generalized the Einstein equations by substituting the Ricci's scalar $R$ by an arbitrary function of $R$, named  $f(R)$ theory of gravity. Another theories, view as alternative to the GR, also have been undertaken, still in the way to better explain the obscure content of the universe, as, $f(R,G)$\cite{eti21}-\cite{eti22}, $f(R,T)$\cite{eti23}-\cite{eti24} and $ f (R,T, R_{\mu\nu }T^{\mu\nu})$\cite{eti25}-\cite{eti26}, where  $G$, $T$, $R_{\mu\nu}$ and $T_{\mu\nu}$ are the invariant of Gauss-Bonnet, the trace of energy-momentum tensor, the Ricci's tensor and the energy-momentum tensor corresponding to the ordinary content of the universe.\par  
In this paper, we focus our attention on $f(R,T)$ theory of gravity. This 
theory has been considered first by  Harko and collaborators  \cite {baf7}. Another authors also have considered this present theory and important results have been found \cite{baf8}-\cite{baf16}.\par
However, any one of the works developed in these papers does not solve the coincidence problem, that is, how does the universe transits from the decelerated phase to the current accelerated phase? \cite{baf18}. The universe is accelerated for the low values of the redshift  and the one that splits the two phases (transition redshift), denoted $z_t$, those the current value is $z_t = 0.46  \pm 0.13 $ \cite{baf19}, \cite{baf20}. The $f(R,T)$ theory can successfully predict the transition from the matter dominated phase to the current accelerated one and several works have described this transition \cite{baf21}. The reconstruction of $f(R,T)$ models describing the matter dominated and accelerated phases had been performed in \cite{baf17}.
Various works also have been performed, still in the optic to better explore this transition with interesting results( see  \cite{baf22}, \cite{baf23} and \cite{eti6}).\par
In this paper, we focus our attention on the epoch after the recombination where, beside the well known ordinary matter (the dust), there is the self-interacting matter, called collisional matter. The model of collisional matter has been studied in some works  within others theories of gravity, leading to interesting results \cite{eti1},\cite{eti7}, \cite{eti8}, \cite{eti9}. This approach of considering new form of matter, instead of the cold dark matter can edify us on the choice of the models of modified gravity.  Oikonomou and collaborators \cite{eti4} have studied the time evolution of the cosmological parameters during the late time acceleration of the universe with the presence of the collisional matter in the framework of modified $f(R)$ gravity. In this paper, we have extended the same idea to the $f(R,T)$. Some $f(R,T)$ models have been considered and the behaviours of the cosmological parameters have been performed and compared with the $\Lambda$CDM model. We see within many results that depending on the input parameters according to the model under consideration, the inclusion of collisional matter may lead to a better explanation of the phase transition, comparatively to the model where just the usual ordinary matter is considered. 
\par
The paper is organized as follows: In section \ref{sec2} we describe the general formalism of $f(R,T)$ theory of gravity. The collisional matter that self-interacts is presented in the section \ref{sec3}. The section \ref{sec4} is devoted to the study of time evolution of the cosmological parameters where the universe is considered to be filled by the usual ordinary matter and the collisional one. Here, for the study of these cosmological parameters, we focus our attention on the transition from the decelerated phase to the accelerated one. In the section \ref{sec5} we examine the  evolution of the equation of state of the dark energy where the matter content is assumed as a fluid is composed by the collisional matter and the radiation. The conclusion and perspectives are presented in the section \ref{sec6}.


\section{General formalism in $f(R,T)$ gravity}\label{sec2}

In this section we present the generality of $f(R,T)$ theory  by substituting the curvature scalar $R$ of the GR by a function of $R$ and the trace $T$, and writing the action as \cite{eti5}
\begin{eqnarray}
S =  \int \sqrt{-g} d^{4}x \Big[\frac{1}{2\kappa^2} f(R,T)+\mathcal{L}_m \Big]\,\,, \label{1}
\end{eqnarray}
where $R$, $T$ denote the curvature scalar and the trace of the energy-momentum tensor, respectively, and $\kappa^2=8\pi \mathcal{G}$, $\mathcal{G}$ being the gravitation constant.\par
The energy-momentum tensor associated to the matter is defined by
\begin{eqnarray}
T_{\mu\nu}=-\frac{2}{\sqrt{-g}}\frac{\delta\left(\sqrt{-g}\mathcal{L}_m\right)}{\delta g^{\mu\nu}}.\label{2}
\end{eqnarray} 
Let us assume that the matter Lagrangian density $L_m$ only depends on the components of the metric tensor, but not on its derivatives. Thereby, one gets  
\begin{eqnarray}
T_{\mu\nu} = g_{\mu\nu}L_{m}-\frac{{2}{\partial{L_{m}}}}{\partial{g^{\mu\nu}}}. \label{3}
\end{eqnarray}
Within the metric formalism, varying the action $(\ref{1})$ with respect to the metric, one obtains the following field equations  \cite{eti5}
\begin{eqnarray}
f_{R}R_{\mu\nu}-\frac{1}{2} g_{\mu\nu}f(R,T)+(g_{\mu\nu}\Box-\nabla_{\mu}\nabla_{\nu})f_{R}= \kappa^{2}T_{\mu\nu}-
f_{T}(T_{\mu\nu}+\Theta_{\mu\nu})\,,\label{4}
\end{eqnarray}
where $f_{R}, f_{T}$  are the partial derivatives of $ f(R,T) $ with respect
to $ R $and $T $ respectively. The tensor $\Theta_{\mu\nu}$ is determined by 
\begin{eqnarray}
\Theta_{\mu\nu}\equiv g^{\alpha\beta}\frac{\delta T_{\alpha \beta}}{\delta g^{\mu\nu}}=-2T_{\mu\nu}+g_{\mu\nu}\mathcal{L}_m
-2g^{\alpha\beta}\frac{\partial^2 \mathcal{L}_m}{\partial g^{\mu\nu}\partial g^{\alpha \beta}}\label{5}.
\end{eqnarray}
As mentioned  in our introduction, we assume that the whole content of the universe is a perfect fluid.  Then, by setting the matter Lagrangian density to $\mathcal{L}_m=-p_m$, the energy-momentum tensor may take the following expression (see appendix)
\begin{eqnarray}
T_{\mu\nu}=\left(\rho_m+p_m \right)u_{\mu}u_{\nu}-p_m g_{\mu\nu} \label{6},
\end{eqnarray}
where the four-velocity satisfies the relations $ u_\mu u^{\mu} = 1$ and $u^{\mu}\nabla_\nu u_\mu $ =0. {\bf Note that the expression (\ref{6}) is obtained with the consideration according to what the pressure does not depend on the metric tensor. Within this consideration according to what the Lagrangian density does not depend of the metric tensor, the contribution of the last term of $(\ref{5})$ vanishes and this equation takes the following form (see also Eq. (21) of \cite{baf7} )
\begin{eqnarray}
\Theta_{\mu\nu}= -2T_{\mu\nu} -p_m g_{\mu\nu}.\label{7}
\end{eqnarray}
However, in the case of the electromagnetic field or scalar field, there is actively contribution of the last term of (\ref{5}) because in these cases the respective Lagrangian densities $\mathcal{L}=g^{\alpha\gamma}g{\beta\sigma}F_{\alpha\beta}F_{\gamma\sigma}$ and  $\mathcal{L}=g^{\mu\nu}\partial_{\mu}\phi\partial_{\nu}\phi$, explicitly depend on the metric. (See also comments after Eq.(18) and Eq.(19) of \cite{baf7}).}
\par
Making use of Eq (\ref{7}), the field equation (\ref{4}) can be reformulated as 
\begin{eqnarray}
R_{\mu\nu}-\frac{1}{2}Rg_{\mu\nu} = \kappa_{eff}^{2} T^{eff}_{\mu\nu}, \label{8}
\end{eqnarray} 
where $ \kappa_{eff}^{2}= \frac{\kappa^2 +f_T}{f_R} $ is the effective gravitational constant  and
\begin{eqnarray}
 T^{eff}_{\mu\nu} = \Bigg[T_{\mu\nu}+\frac{1}{\kappa^2+f_T}\bigg( \frac{1}{2}g_{\mu\nu}(f-Rf_R)+ f_Tp_m g_{\mu\nu}-(g_{\mu\nu}\square-\nabla_\mu\nabla_\nu) f_R \Bigg) \Bigg], \label{9}
\end{eqnarray}
representing the effective energy-momentum tensor of the matter.\par

Taking into account the covariant divergence of Eq.(\ref{8}), with we use the identities
  $  \nabla^ {\mu} (R_{\mu\nu}-\frac{1}{2}Rg_{\mu\nu})=0  $,  and 
$  \left(g_{\mu\nu}\nabla^{\mu}\Box-\nabla^{\mu}\nabla_{\mu} \nabla_{\nu}\right)f_{R}= -R_{\mu\nu}\nabla^{\mu}f_{R} $,  on gets 
 
\begin{eqnarray} 
  \nabla^{\mu}T_{\mu\nu}= 
 \frac{1}{\kappa^2-f_T}
 \left[f_T\nabla^{\mu}\Theta_{\mu\nu}+\left(T_{\mu\nu}
+\Theta_{\mu\nu}\right)\nabla^{\mu}f_T-
 \frac{1}{2}g_{\mu\nu}f_{T}\nabla^{\mu}T\right].
 \label{222}
\end{eqnarray}
 After some elementary 
transformations, Eq.~(\ref{222}) takes the following expression 
\begin{eqnarray}  
 \nabla_{\mu}T^{\mu}_{\nu}= \frac{1}{\kappa^2+f_T}
 \Bigg\{-\omega\delta^{\mu}_{\nu}f_T
 \partial_{\mu}\rho  +  (1-3\omega)  
 \Bigg[
 \Big(-T^{\mu}_{\nu}- \omega \rho \delta^{\mu}_{\nu}
 \Big)f_{TT}-
 \frac{1}{2}\delta^{\mu}_{\nu}f_T \Bigg]
 \partial_{\mu}\rho\Bigg\}.
 \label{223}
 \end{eqnarray}

  It comes from the above expression that for the $f(R, T)$ theory of gravity, the conservation of the energy-momentum tensor is not automatic. In order the reach this condition, it is important to constraint the second hand of this equation to zero (see Eq.~(\ref{15})). This condition is extremely crucial for determining the suitable expression of the Trace depending part of the algebraic action. {\bf The reference \cite{alvarenga3}  considers the constraint of the conservation to energy-momentum tensor (as in (\ref{15})) and one obtains, after the integration on the right side of (\ref{223}), $f(T)=\alpha_0 T^{(1+3\omega)/[2(1+\omega)]}+\beta_0$, with $\alpha_0,\beta_0$ real constants; (see (19) of \cite{alvarenga3}).}

Here we are interested to the spatially flat Friedmann-Robertson-Walker (FRW) metric 
\begin{eqnarray}
 ds^{2}= dt^{2}-a(t)^{2}[dx^{2}+ dy^{2}+dz^{2}].\label{10}
\end{eqnarray}
Thus, the Ricci scalar in this background is given by
\begin{eqnarray}
R= -6( 2H^2+\dot{H}). \label{11}
\end{eqnarray}

\section{ collisional matter model within $f(R,T)$ theory }\label{sec3} 

This king of matter has been introduced firstly in \cite{eti1,eti3} in the framework of GR and $f(R)$. The evolution of the collisional matter depends on the source that drives the whole content (matter and energy) of the univers.\par 
The study of the model of collisional matter with perfect fluid has been performed widely in \cite{eti4}.
In this work, we recall that our basic assumption is that  the matter has a total mass-energy density, denoted by $ \varepsilon_m $, which is assumed to be depending on two contributions as 
\begin{eqnarray}
\varepsilon_m = \rho_m+ \rho_m \varPi. \label{11}
\end{eqnarray}
Here, $\rho_m$ refers to the part that does not change due to the fact of characterising the usual matter content \cite {eti1},\cite{eti3}. Concerning the term $\rho_m\varPi$, it expresses the energy density part of the energy momentum tensor associated  
with thermodynamical content of the collisional matter. This fluid could not be dust, but possesses a positive pressure and satisfying the following equation of state
\begin{eqnarray}
P_m = w\rho_m,    \label{12} 
\end{eqnarray}
where $w$ denotes the parameter of equation of state of the collisional matter whose values is such that $ 0 < w < 1$ .
The potential energy density is assumed to have the form \cite {eti1},\cite{eti3} ,
\begin{eqnarray}
\varPi = \varPi_0 + w\ln (\frac{\rho_m}{\rho_{m0}}).  \label{13} 
\end{eqnarray}
where the constants $  \rho_{m0}$ and $ \varPi_0$ are their current values.\par
Accordingly, after elementary transformation the total-energy density of the universe can be written as 
\begin{eqnarray}
 \varepsilon_m = \rho_m \bigg(1+ \varPi_0+w\ln (\frac{\rho_m}{\rho_{m0}})\bigg). \label{14} 
\end{eqnarray}
Through the continuity equation, the motions of the volume elements in the interior of a continuous medium can be traduced
\begin{eqnarray}
\nabla^ {\nu} T_{\mu\nu} = 0, \label{15}  
\end{eqnarray}
and the energy-momentum tensor takes the following form 
\begin{eqnarray}
 T_{\mu\nu} = (\varepsilon_m+p_m) u_\mu u_\nu - p_m g_{\mu\nu}, \label{16} 
\end{eqnarray}
where $u_\mu= {dx_\mu}/ {ds}$ is the four velocity, satisfying the relation $ u_\mu u_\nu = 1$. We note here that  $ p_m = P_m $  because the pressure of ordinary matter is negligible.  Making use of the FRW line element (\ref{10}), the conservation law of the Equation (\ref{15}) 
\begin{eqnarray}
\dot{\varepsilon_m}+3\frac{\dot{a}}{a}(\varepsilon_m+p_m)=0,\label{17}
\end{eqnarray}
which, upon consideration of the equations (\ref{12}) and (\ref{14}), leads to \cite{eti1}
\begin{eqnarray}
\rho_m = \rho_{m0} \Bigg(\frac{a_0}{a}\Bigg)^{3}, \label{18} 
\end{eqnarray}
where $ a_0$ is the current scale factor.\par
We can describe the collisional matter by equations (\ref{14}) and (\ref{18}) and use therm in the rest of the manuscript in the next section. The value of  $ \varPi_0$ is equal to \cite{eti1}
\begin{eqnarray}
\varPi_0 = \Bigg(\frac{1}{\Omega_M}-1\Bigg).  
\end{eqnarray}
We will use the same value \cite{eti1},\cite{eti2}  of $\varPi_0= 2.58423$ for making the numerical study.   According to the fact that 
$\rho(t_0)=\rho_0$, and making use of Eq.~(\ref{13}), one gets $\varPi(t_0)=\varPi_0$, meaning that this latter is the current value of the potential energy per unit rest-mass,
associated with the infinitesimal deformations (expansions
or/and compressions) of the fluid (see detail in \cite{eti3}).

\section{Late Time Cosmological evolution in $f(R,T)$ theory}\label{sec4}

\subsection{Deceleration Parameter}

In this subsection we  examine the deceleration parameter $q(z)$ in $f(R,T)$ considering a that the universe, beside the ordinary matter and dark energy, also is filled with collisional matter at late time cosmological evolution. In this optic, we rewritte the field equation (\ref{4}) in the following form
\begin{eqnarray}
3H^2= \frac{1+f_T}{f_R}\varepsilon_m+ \frac{1}{f_R}\bigg[ \frac{1}{2}(f-Rf_R)-3\dot{R}Hf_{RR}+ p_m f_T \bigg], \label{19} 
\end{eqnarray}
\begin{eqnarray}
-2\dot{H}-3H^2 =\frac{1+f_T}{f_R}p_m + \frac{1}{f_R}\bigg[2H\dot{R}f_{RR}+\ddot{R} f_{RR}+ \dot{R}^{2}f_{RRR}-
\frac{1}{2}(f-Rf_R)-  p_m f_T \bigg ], \label{20}
\end{eqnarray}
where we have assumed $ \kappa^2 = 1$. The parameter $H = \frac{\dot{a}}{a}$ denotes the Hubble parameter, the ``dot", the time derivative of $ \partial / \partial_t $. We propose to reformulate the right side of (\ref{19}) and (\ref{20}) respectively in term of effective energy density $\rho_{eff}$ and
pressure $ p_{eff}$, and we write (\ref{19}) and (\ref{20}) as
\begin{eqnarray}
3H^2= \kappa_{eff}^2 \rho_{eff}, \label{21}
\end{eqnarray}
\begin{eqnarray}
-2\dot{H}-3H^2 = \kappa_{eff}^2 p_{eff}. \label{22} 
\end{eqnarray}
Here, the effective energy density $ \rho_{eff}$ and effective pressure $p_{eff}$ are respectively defined as
\begin{eqnarray}
\rho_{eff} = \varepsilon_m+ \frac{1}{1+f_T}\bigg[ \frac{1}{2}(f-Rf_R)-3\dot{R}Hf_{RR}+ p_m f_T \bigg], \label{23}
\end{eqnarray}
\begin{eqnarray}
p_{eff} = p_m + \frac{1}{1+f_T}\bigg[2H\dot{R}f_{RR}+ \ddot{R} f_{RR}+\dot{R}^{2}f_{RRR}-\frac{1}{2}(f-Rf_R)-  p_m f_T \bigg ]  \label{24}
\end{eqnarray}
In this way, we have a matter fluid representation of the so-called geometrical dark energy in $ f(R,T)$ gravity with the energy
density $ \rho_{DE} = \rho_{eff} -\varepsilon_m  $ and pressure $ p_{DE} = p_{eff} - p_m$ .\par
From the conservation law, the effective energy density evolves as
\begin{eqnarray}
\frac{d(\kappa_{eff}^2 \rho_{eff})}{dt}+3H\kappa_{eff}^2( \rho_{eff}+p_{eff})=0 \label{25}
\end{eqnarray}
By making use of the equations (\ref{19}) and (\ref{20}), (\ref{25}) can be rewritten as
\begin{eqnarray}
18 \frac{f_{RR}}{f_R}H(\ddot{H}+4H\dot{H})+3(\dot{H}+H^2)+\frac{1+f_T}{f_R}\varepsilon_m+ p_m \frac{f_T}{f_R}+\frac{f}{2f_R}=0. \label{26} 
\end{eqnarray}
One can  rewrite the above equation using the redshift $ z = \frac{1}{a}-1$ as follows
\begin{eqnarray}
\frac{d^2{H}}{dz^2}= \frac{3}{1+z} \frac{dH}{dz}-\frac{1}{H} {\bigg(\frac{dH}{dz}\bigg)}^{2}-
\frac{3f_R \bigg( H^2-(1+z)H\frac{dH}{dz}\bigg)+\frac{f}{2}+p_m f_T+(1+f_T)\varepsilon_m }{18H^3 f_{RR} {(1+z)}^2}. \label{27}
\end{eqnarray}
As convention in this paper, we set to unity the current value of the scale factor.
We also assume the algebraic function as a sum of two independent functions, such that, $ f(R,T) = f(R)+f(T)$.
The  $ f(T)$ model is the one obtained by imposing the conservation of the energy-momentum tensor {\bf (\ref{223})}.

The dynamics and stability of this model are studied in \cite{eti5} and interesting results have been found.
This model reads
\begin{eqnarray}
f(T) \propto  T^{\alpha}, \label{000}
\end{eqnarray}
with $\alpha = \frac{1+3w}{2(1+w)}$.
Here the trace of the energy- momentum tensor (\ref{16}) reads $ T= \varepsilon_m-3p_m.$ \par
From Equations (\ref{14}) and (\ref{18}), the model $ f(T)$ and its first derivative are
\begin{eqnarray}
f(T) = {(1+z)}^{3\alpha} {\rho_{m0} }^{\alpha}{\bigg[1+\varPi_0-3w(1-\ln(1+z)) \bigg]}^{\alpha}, \label{28}  
\end{eqnarray}
\begin{eqnarray}
f_T = \alpha {(1+z)}^{3(\alpha-1)} {\rho_{m0} }^{(\alpha-1)}{\bigg[1+\varPi_0-3w(1-\ln(1+z)) \bigg]}^{\alpha-1}. \label{29}  
\end{eqnarray}
By using (\ref{28}) and (\ref{29}), one rewrite (\ref{27}) as
\begin{eqnarray}
\frac{d^2{H}}{dz^2}= \frac{3}{1+z} \frac{dH}{dz}-\frac{1}{H} {\bigg(\frac{dH}{dz}\bigg)}^{2}-J_1(z)-J_2(z)-J_3 (z), \label{30} 
\end{eqnarray}
where 
\begin{eqnarray}
J_1 (z) =  \frac{3f_R\bigg(H^2-(1+z)H\frac{dH}{dz}\bigg)+\frac{f(R)}{2}} {18H^3f_{RR}{(1+z)}^{2}},\label{31} 
\end{eqnarray}
\begin{eqnarray}
J_2 (z) = \frac{\rho_{m0}(1+z)\bigg(1+\varPi_0+3w\ln(1+z)\bigg)}{18H^3 f_{RR}}, \label{32} 
\end{eqnarray}
\begin{eqnarray}
J_3 (z) =\bigg(\alpha(1+w)+\frac{1}{2}(1-3w) \bigg) {\rho_{m0}}^{\alpha-1}{(1+z)}^{3\alpha-1}
\bigg{(1+\varPi_0-3w(1-\ln(1+z))\bigg)}^{\alpha-1} J_2(z) \label{33} 
\end{eqnarray}
The equation (\ref{30}) is the main tool to be used for analysing the cosmological evolution of the considered $f(R,T)$ models. The numerical analysis also will be performed in order to find the Hubble parameter. In a first step, by using the expression 
\begin{eqnarray}
q(z) =\frac{(1+z)}{H(z)}\frac{dH}{dz}-1,  \label{33}  
\end{eqnarray}
we can perform the study of the deceleration parameter with the account of the collisional matter, this latter assumed as a function of the redshift $z$.  In this rubric we will focus our attention on the transition from the decelerated phase to the accelerated one. Therefore, we will compare the results with the case where the non-collisional matter (pressure-less) is considered, and also compare with the  $ \Lambda CDM $ case. \par
We recall that in  \cite{eti2}, the corresponding deceleration parameter of the expansion in the flat friedmann cosmology for the $ \Lambda CDM $ model is given by
\begin{eqnarray}
q(z) = \Bigg[ \frac{\Omega_{m0}}{2}{(1+z)}^{3}-L_0\Bigg] {\Bigg [\Omega_{m0}{(1+z)}^{3}+L_0 \Bigg]}^{-1}, \label{33.1}
\end{eqnarray}
where  $\Omega_{m0}=0.279 $ and $ L_0 = 0.721$ \cite{eti2}.\par
Moreover,  the  effective equation of state parameter $ w_{eff}$ is defined as,
\begin{eqnarray}
 w_{eff} &=& \frac{p_{eff}}{\rho_{eff}}  \\
 &=& -1+\frac{2(1+z)}{3H(z)} \frac{dH}{dz},   \label{34}
\end{eqnarray}
and we present its evolution versus the redshift. A comparison with the results to be found will be done with the results coming from pure $f(R,T)$ models with non-collisional matter.

\subsection{ Study of particular cases $f(R)$ models in modified  $f(R,T)$ gravity universe with collisional matter  }

In this subsection we discuss two $f(R)$ models, performed in \cite{eti4,eti6} for describing the late time cosmological evolution of our universe under the consideration that it is filled with collisional (self-interacting) matter.  Thereby, we solve numerically the  differential equation about $H(z)$.\par
The task here is to find the models that should be in agreement with the observational data, namely the deceleration parameter $q(z)$ and the parameter of effective equation of state $w_{eff}$.\par
The first model  considered is a modified power-law model of the curvature scalar 
\begin{eqnarray}
f(R) = \lambda_{0} {(\lambda +R)}^{n}; \label{00}
\end{eqnarray}
where $\lambda_{0}$, $\lambda$ and $ n$ are positive constants. \par 
The second, is an exponential model 
\begin{eqnarray}
f(R)=  R_0  e ^{\beta R}, \label{001}
\end{eqnarray}
where $R_0$ and $\beta$  are constant parameters.\par
Numerical plots of the deceleration parameter $q(z)$ and the effective equation of state $w_{eff}$  are presented from (\ref{30}) with collisional matter $(w =0.6)$ and non-collisional matter $(w=0)$ for each model.

\begin{figure}[h]
\centering
\begin{tabular}{rl}
\includegraphics[width=7cm, height=7cm]{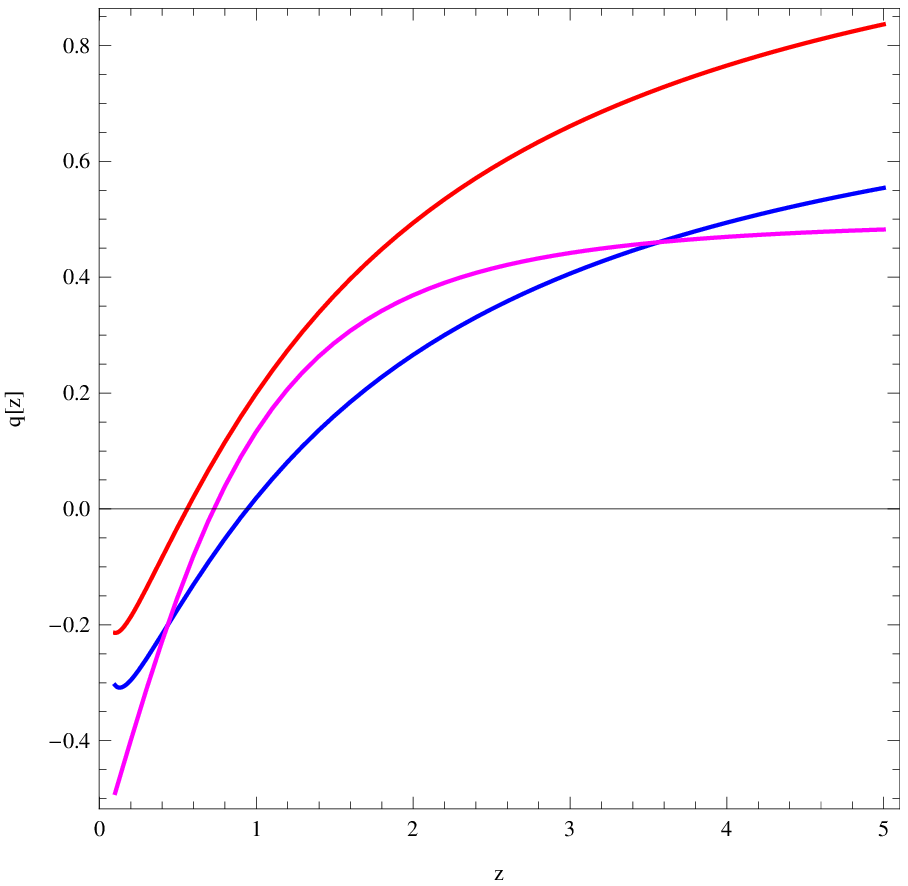}&
\includegraphics[width=7cm, height=7cm]{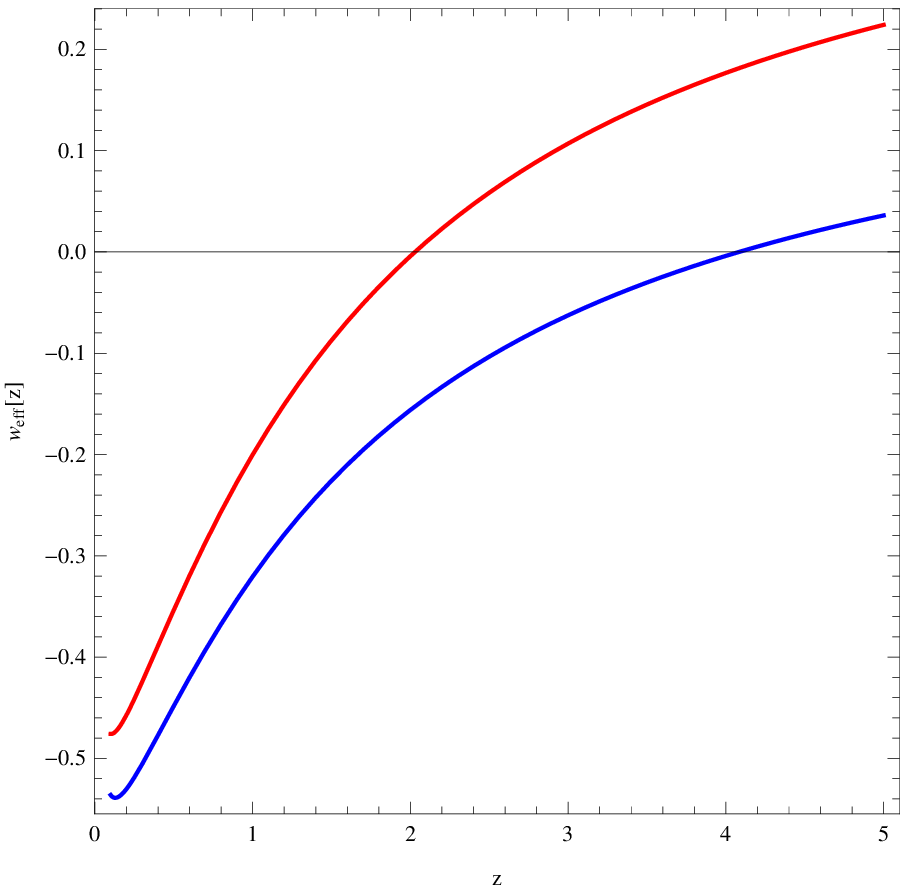}
\end{tabular}
\caption{The graphs show the evolution of $q(z)$ and $ w_{eff}$ versus z for the model $   f(R,T)=\lambda_{0} {(\lambda +R)}^{n}+T^{\alpha} $. The graphs are plotted for $\lambda_{0}=1$, $\lambda 
=13.5$ and $ n=0.5$ The red, blue and the magenta refer to non-collisional, collisional matter and $\Lambda CDM$ model, respectively.}
\label{fig1}
\end{figure}

 \begin{figure}[h]
 \centering
 \begin{tabular}{rl}
 \includegraphics[width=7cm, height=7cm]{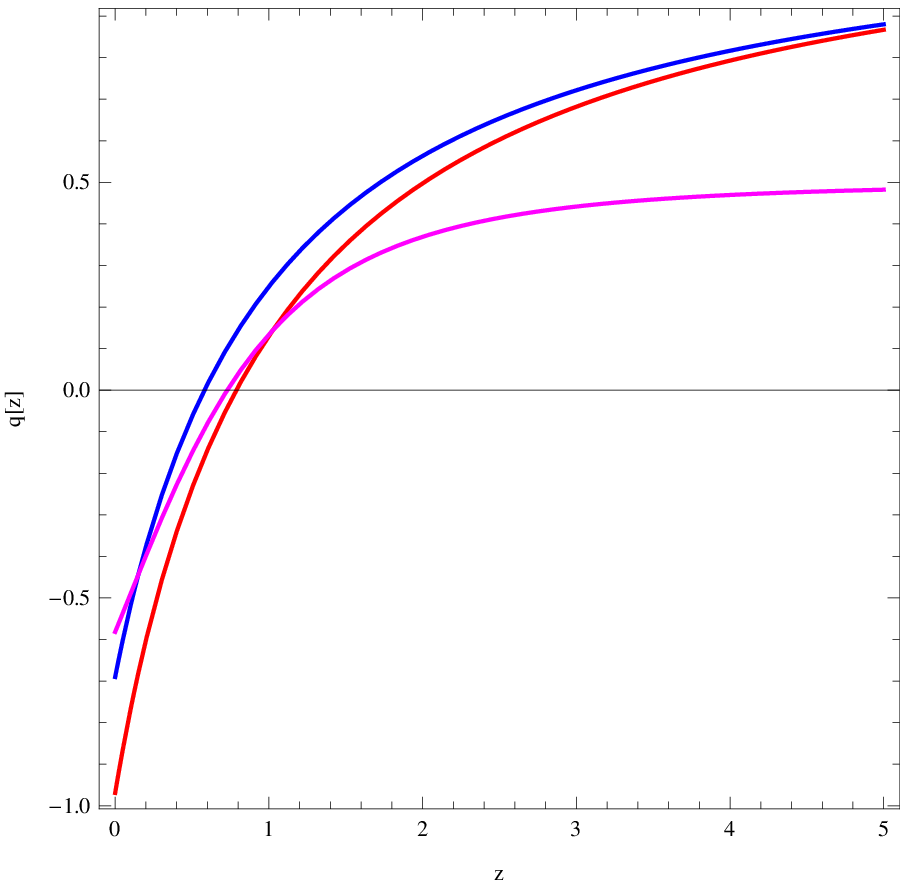}&
 \includegraphics[width=7cm, height=7cm]{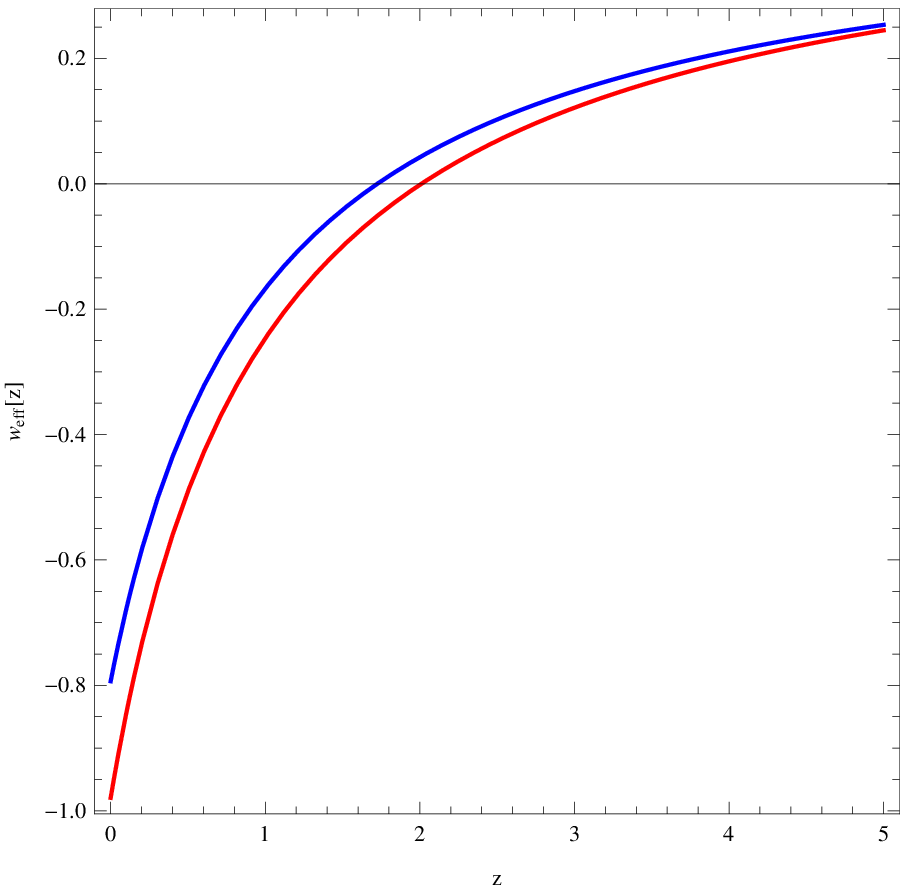}
 \end{tabular}
 \caption{The graph shows the evolution of $q(z)$ and $ w_{eff}$ versus z for the model $ f(R,T)= R_0 e^{\beta   R}+T^{\alpha}$.  
The graph are plotted for $R_0=1$ and $\beta=1.5.$ The red, blue and the magenta refer to non-collisional matter, collisional
matter and $\Lambda CDM$ model, respectively.}
\label{fig2}
\end{figure}

For Fig.~$1$, related to the first $f(R,T)$ model, we see that, at the left hand side, traducing the evolution of 
the parameter of deceleration the curve representative of the collisional matter is nearby that of $\Lambda$CDM than
the curve representative of the non-collisional matter.  We also see that, in the order, collisional matter, $\Lambda$CDM 
and non-collisional matter, the transition from the decelerated phase to the accelerated phase is realized from high to the
low value of the redshift. Concerning the curve representative of $w_{eff}$ the same behaviours of the phase transition order is confirmed.  We see that the behavior of the cosmological evolution of the deceleration parameter $q(z)$ and the effective
equation of state parameter $ w_{eff}$ for the model $ f(R,T)= \lambda_{0} {(\lambda +R)}^{n}+T^{\alpha}$ with and without collisional matter are similar
to the model  $ f(R) =\lambda_{0} {(\lambda +R)}^{n}$ without collisional matter obtained by the authors of (\cite{eti4}).
Moreover, as the same authors, we see that the transition from the decelerated phase to the accelerated phase is realized from
high to the low value of the redshift in same order of the content of universe. Also, we remark that, the transition of the redshift
obtained  by this authors for $f(R)$ model without collisional matter is equal to the one obtained for $f(R,T)$ model without collisional
matter, whereas the transition of the redshift  to the $f(R)$ model with collisional matter is strongly than the transition
of the redshift for  $f(R,T)$ model. For this observation, we can conclude that the depending of the trace $T$ 
 of the energy-momentum tensor in this model
is very important to reproduce the $\Lambda$CDM model.

\par
Regarding the second model, whose cosmological parameters are plotted at Fig.~$2$, we see that for  the transition from the
decelerated to the accelerated phases, from the high to low redshifts, the order is non-collisional matter, $\Lambda$CDM and 
collisional matter. Moreover, we see that in the accelerated phase, the collisional matter is more nearby the $\Lambda$CDM than the non-collisional one, while in the decelerated phase, both the collisional and non-collisional matters, almost follow the same trajectory for high redshifts.
This aspect is very important because revels that the collisional takes its origin from the ordinary matter.
We then conclude that at early times there is no collisional matter, but just non-collisional, and as the time evolves, the collisional matter start being created from the non-collisional matter.
After crossing the transition line, the collisional matter approaches more the $\Lambda$CDM as the low redshifts are being reached. We see that the behaviour of the cosmological evolution of the deceleration parameter $q(z)$ and the effective
equation of state parameter $ w_{eff}$ for the model $ f(R,T)= R_0  e ^{\beta R}+T^{\alpha}$ with and without collisional matter is similar
to the model  $ f(R) = R_0  e ^{\beta R} $ with and without collisional matter obtained by the authors of (\cite{eti4}).
At the high value of the redshifts, the trajectory of this parameters is bewildered, both the collisional and non-collisional
matter for  $ f(R)$ and  $ f(R,T)$ models

\subsection{ $f(R,T)$ Models with Cardassian Matter}

The cardassian model is firstly developed by  \cite{eti7},\cite{eti8},\cite{eti9} and is another approach to describe the self interacting  matter. According to the observational  it is well known that the universe is flat and accelerating; the cardassian model of matter is characterised by negative pressure and there is no vacuum energy whatsoever. In this subsection we perform the same study as in the previous section and use the models  $f(R)$ models (\ref{00}), (\ref{001}).
 According to the so-called Cardassian model of matter \cite{eti7},\cite{eti8},\cite{eti9}, the total energy density $\varepsilon_m$ of matter is defined by 
 \begin{eqnarray}
 \varepsilon_m = \rho + \rho K(\rho) \label{35}
 \end{eqnarray}
 where $\rho $ is related to the ordinary matter-energy density and  $\rho K (\rho)$ describes the cardassian model of matter and is general, a  function of the ordinary mass-energy density $\rho$.\par
According to the original Cardassian model, one has 
\begin{eqnarray}
\rho K(\rho) = B {\rho} ^{n'}, \label{36}
\end{eqnarray}
where $B$ is a real number and  it is fixed $ n' < \frac{2}{3}$ in order to guarantee the acceleration.\par
By taking into account the relation (\ref{36}), we reobtain the total energy density of matter (\ref{35}) as,
\begin{eqnarray}
\varepsilon_m = \rho + B {\rho} ^{n'}. \label{37}
\end{eqnarray}
 It is important to point out here that the above expression may be expressed in term of the redshift $z$. To do so we take into account (\ref{18}) and cast it on the form $\rho =\rho_{m0}{(1+z)}^{3}$, where we assumed the current scalar factor $a_0=1$. Therefore, the equation (\ref{37}) becomes
\begin{eqnarray}
\varepsilon_m =\rho_{m0}{(1+z)}^{3}\Biggr(1+B{\rho_{m0}}^{(n'-1)}{(1+z)}^{3(n'-1)} \Biggl)
\end{eqnarray}

Following the same idea as \cite{eti4},  we assume that the late time evolution is governed by the geometric dark fluid with negative pressure  coming from the $f(R,T)$ models plus a gravitating fluid of positive pressure, satisfying
the following equation of state
\begin{eqnarray}
p = w_k \rho.\label{38} 
\end{eqnarray}
We mention in this model that the pressure has to be assumed as negative quantity, $p<0$, because the cardassian model of matter possesses negative pressure. In such a situation, this negative has to be assumed as the one which generates the acceleration of our universe. \par
Adopting the assumption that the total energy density of matter is (\ref{37}), the main equation (\ref{27}) can be used for analysing the evolution of the viable $f(R,T)$ models considered with Cardassian self-interacting matter, yielding 
\begin{eqnarray}
\frac{d^2{H}}{dz^2}= \frac{3}{1+z} \frac{dH}{dz}-\frac{1}{H} {\bigg(\frac{dH}{dz}\bigg)}^{2}-J'_1(z)-J'_2(z)-J'_3 (z), \label{39} 
\end{eqnarray}
where
\begin{eqnarray}
J'_1 (z)= \frac{3f_R\bigg(H^2-(1+z)H\frac{dH}{dz}\bigg)+\frac{f(R)}{2}} {18H^3f_{RR}{(1+z)}^{2}}, \label{01} 
\end{eqnarray}

\begin{eqnarray}
J'_2 (z)=\frac{\rho_{m0} (1+z)\bigg(1+B{\rho_{m0}}^{(n'-1)} {(1+z)}^{3(n'-1)}\bigg)}{18H^3 f_{RR}}, \label{40} 
\end{eqnarray}
\begin{eqnarray}
J'_3 (z)= {\rho_{m0}}^{\alpha-1}{(1+z)}^{3(\alpha-1)}(\alpha+1/2)(1+w_k) 
{\bigg(1+w_k+B{\rho_{m0}}^{(n'-1)}{(1+z)}^{3(n'-1)}\bigg)}^{\alpha-1} J'_2(z).\label{41} 
\end{eqnarray}
By adopting the same treatment as in the previous section, we present the evolutions of the deceleration parameter $ q(z)$ and the effective equation $ w_{eff}$ of state for the late time cosmological  evolution of the universe with the $f(R,T)$ models including Cardassian Matter.
  
  \begin{figure}[h]
  \centering
  \begin{tabular}{rl}
  \includegraphics[width=7cm, height=7cm]{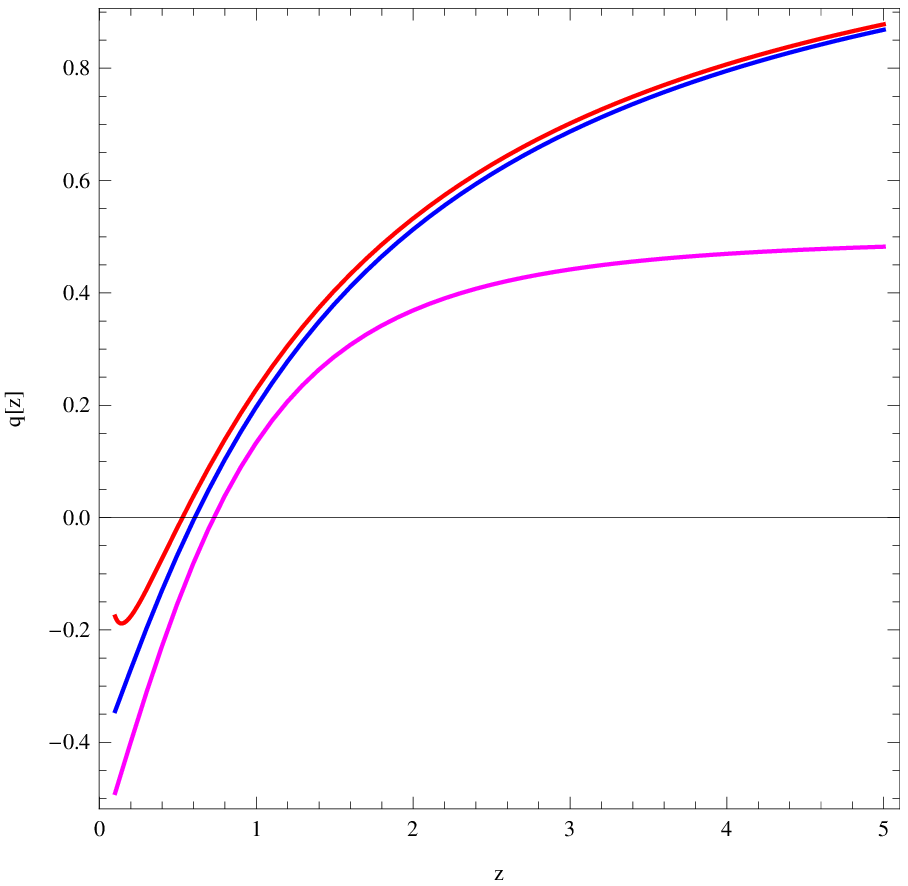}&
  \includegraphics[width=7cm, height=7cm]{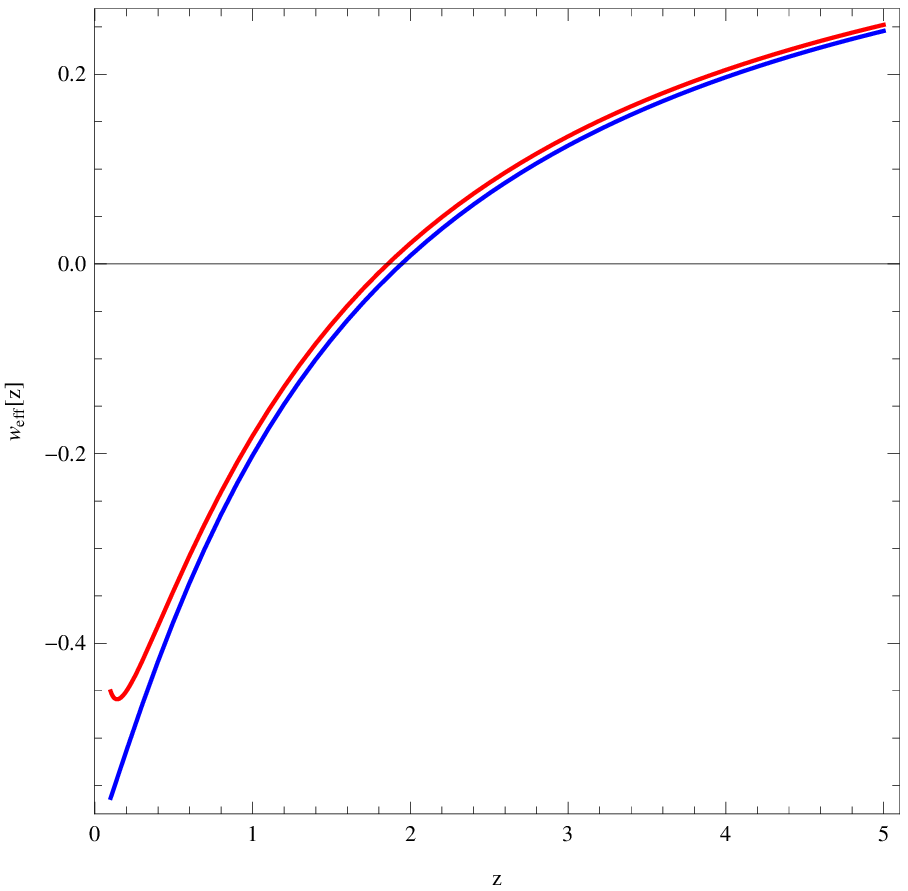}
  \end{tabular}
  \caption{The graph shows the evolution of $q(z)$ and $ w_{eff}$ versus z for the model $ f(R,T)=\lambda_{0} {(\lambda +R)}^{n}+T^{\alpha} $.  
 The graph are plotted for $\lambda_{0}=1$, $\lambda =13.5$ and $ n=0.5$ The red color, blue and the 
  magenta refer to collision-less, cardassian and $\Lambda CDM$ model, respectively and we use $B=0.2$, $n'=-3$.}
 \label{fig3}
  \end{figure}
 
  \begin{figure}[h]
  \centering
  \begin{tabular}{rl}
  \includegraphics[width=7cm, height=7cm]{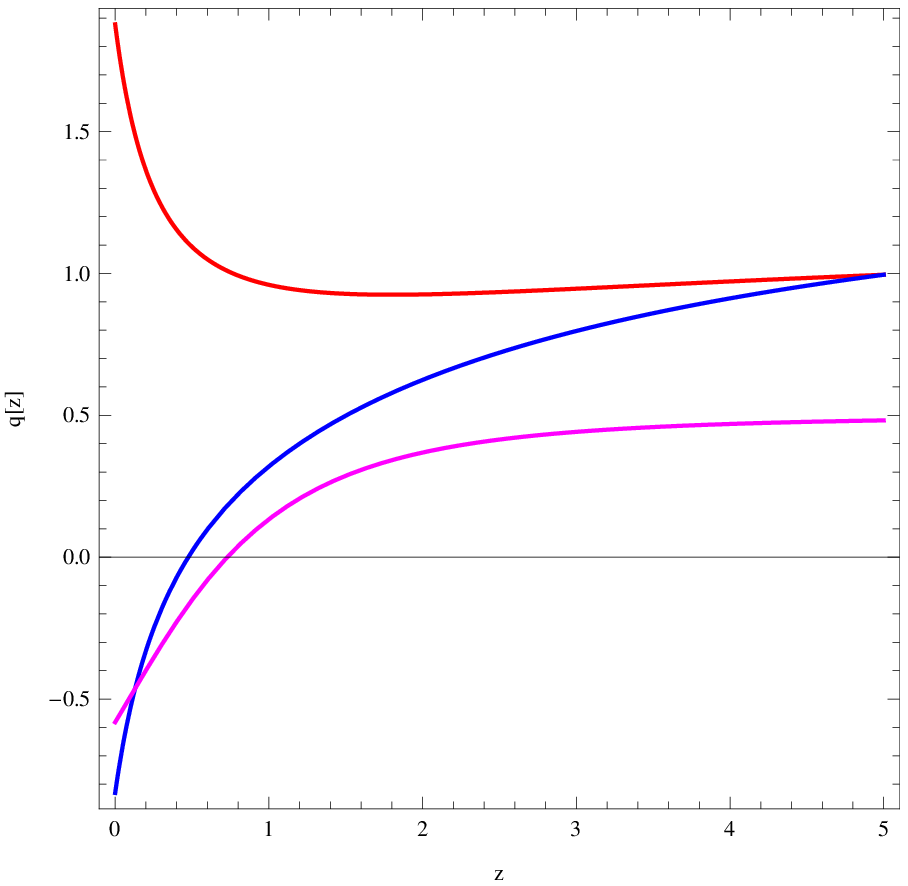}&
  \includegraphics[width=7cm, height=7cm]{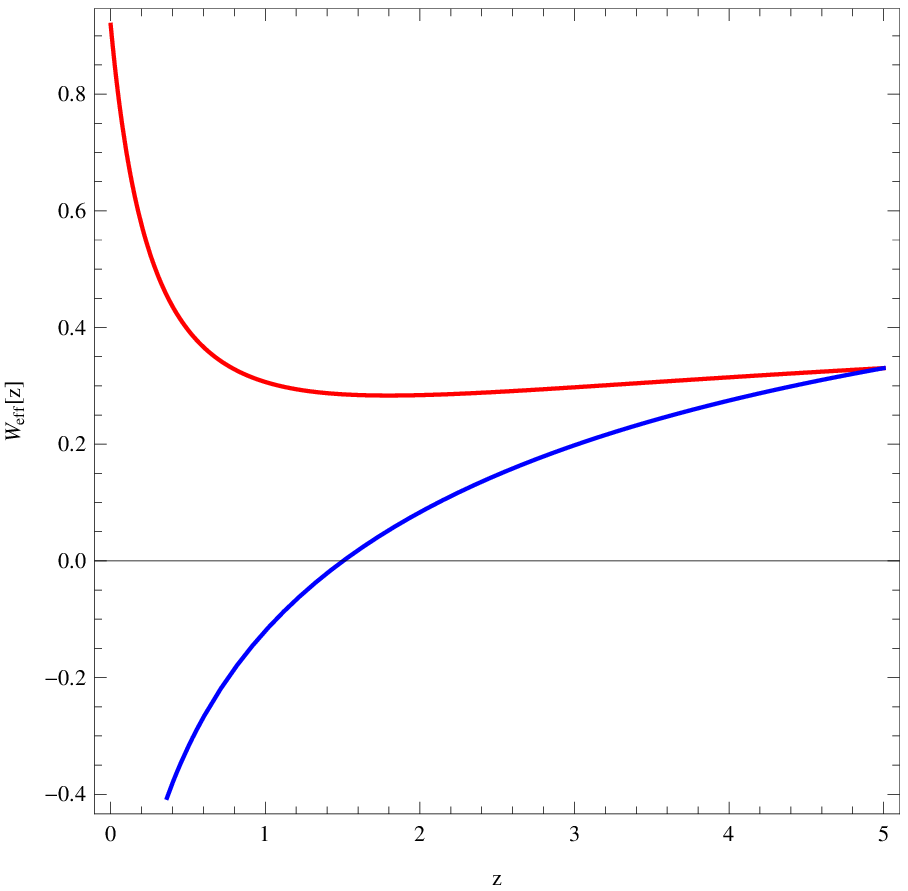}
  \end{tabular}
  \caption{The graphs show the evolution of $q(z)$ and $ w_{eff}$ versus $z$ for the model $ f(R,T)= R_0  e ^{\beta R}+T^{\alpha}$.
  The graphs are plotted for $ R_0=0.5$ and $\beta=1.4$. The red, blue and magenta refer to collision-less matter, cardassian matter and $\Lambda CDM$ model, respectively and we use $B=0.2$, $n'=-3$.}
  \label{fig4} 
\end{figure}
Here, for the first model, Fig.~$3$, we see that the transition from the  phase of deceleration to the acceleration one 
is realized, from the high to low-redshifts, in the order, $\Lambda$CDM, cardassian matter and collision-less matter.
In the decelerated phase, both the cardassian and collision-less matters are confused, while, within the accelerated phase,
the cardassian matter  detached from the collision-less one and start approaching the $\Lambda$CDM model as the low-redshifts are being reached.
 As the authors of (\cite{eti4}), we see, both the cardassian matter and collision-less matter, 
 that the cosmological evolution of the deceleration parameter $q(z)$ and the effective
equation of state parameter $ w_{eff}$ for the models $ f(R,T)= \lambda_{0} {(\lambda +R)}^{n}+T^{\alpha}$  and $ f(R)= \lambda_{0} {(\lambda +R)}^{n}$ are similar behavior and are same
trajectory at high value of the redshifts. The transition of the redshift for each model with cardassian matter and collision-less
matter is inferior at the transition of the $\Lambda CDM$ model.\par 
Concerning the second model, Fig.~$4$, the cardassian and collision-less matter are also confused for high redshifts
and as the low-redshifts are being reached, the cardassian matter approximate the $\Lambda$CDM model.
 We observe at the low value of the redshifts, with cardassian matter, that the behavior of the $ f(R,T)= R_0  e ^{\beta R}+T^{\alpha}$ model
is not similar of the model  $ f(R) = R_0  e ^{\beta R}$ obtained by the authors of (\cite{eti4}), but, when the redshift $z$ increase,
the trajectory of this models with cardassian matter are similar behaviour and bewildered with the collision-less matter trajectory.

 
\section{Equation of state for dark energy with collisional Matter Fluid in $f(R,T)$ gravity}\label{sec5}

The work performed in this section is about the cosmological evolution  of the universe through $f(R,T)$ models in presence of matter composed by collisional matter and radiation.\par
We focus our attention on the exponential $f(R)$ model  \cite{eti10},\cite{eti11} given by
\begin{eqnarray}
f(R)= R-\beta R_S (1-e^{-R/R_S}) \label{aa}, 
\end{eqnarray}
where $ c_1 = -\beta R_S $ and $ c_2 = R_S $ constants with curvature scalar dimension. The conditions for having a viable exponential model has been studied in  \cite{eti12}.\par
Within flat FRW metric and considering the above content of the universe, the field equations, from the general form  (\ref{4}), take the following form 
\begin{eqnarray}
3f_R H^2 = (1+f_T) \rho_ {matt} +\frac{1}{2}(f-R f_{R})-3H \dot{f_R}+P_{matt} f_T, \label{42}
\end{eqnarray}
\begin{eqnarray}
-2f_R \dot{H} = (1+f_T) (\rho_ {matt}+P_{matt})+\ddot{f_R}-H\dot{f_R}, \label{43} 
\end{eqnarray}
where  $\rho_ {matt}$ and $P_{matt}$ denote the energy density and pressure of all perfect fluids of generic matter, respectively. \par
Concerning the component to be considered here, we assume that the universe is filled with collisional matter (self-interacting matter) and the relativistic matter (radiation). This means that in our analysis, the contribution from the radiation may play an important role. Therefore, the matter energy density, $\rho_ {matt}$, is given by
\begin{eqnarray}
\rho_ {matt} = \varepsilon_m +\rho_{r0} a^{-4}, \label{44} 
\end{eqnarray}
where $\rho_{r0}$ is the current energy density of radiation. The pressure of all perfect fluids of matter is given
by 
\begin{eqnarray}
P_{matt} = p_m + p_r. \label{0000}
\end{eqnarray}
The first term characterises the pressure of collisional matter, given by (\ref{12}), and the second, is related to the radiation. Making use of (\ref{14}) and (\ref{18}), one can put Eq. (\ref{44}) into the following form 
\begin{eqnarray}
\rho_ {matt} = \rho_{m0}a^{-3}\bigg(1+\varPi_0+3w \ln(a) \bigg)+ \rho_{r0} a^{-4}. \label{45}
\end{eqnarray}
In the same way, Eq.~(\ref{42}) can be rewritten as 
\begin{eqnarray}
H^2 + H^2 f_{RR} \frac{dR}{d\ln a}-\frac{1}{6}(f-R)+ (1-f_R)(H\frac{dH}{d\ln a}+H^2) = \frac{1}{3}\rho_ {matt}+
\frac{f_T}{3}(\rho_ {matt}+P_{matt}), \label{46} 
\end{eqnarray}
while the scalar curvature R can be expressed as
\begin{eqnarray}
R = - (12H^2+6H\frac{dH}{d\ln a}). \label{47}  
\end{eqnarray}
Through the Eq. (\ref{45}), we assume that the total matter-energy density becomes
\begin{eqnarray}
\rho_ {matt} = \rho_{m0} ( g(a)+ \chi a^ {-4}). \label{48}
\end{eqnarray}
In the previous expression, $\chi$ is defined by $\chi = {\rho_{r0}}/{\rho_{m0}} \simeq 3.1 \times 10^ {-4}$,  
$\rho_{r0}$ being the current energy density of radiation. The parameter $ g(a)$, describing the nature of the collisional matter (view as perfect fluid), is equal  to
\begin{eqnarray}
g(a) = a^{-3} \bigg(1+\varPi_0+3w \ln(a) \bigg). \label{49}      
\end{eqnarray}
Note that for  the non-collisional matter (assumed as the dust), for which the parameter $w=0$, one gets  $ g(a)= a^{-3} $. \par
In the optic to better study the cosmological evolution of the $f(R,T)$ models in the framework of flat FLRW universe, we may introduce the variable   \cite{eti4'}
\begin{eqnarray}
y_H \equiv \frac{\rho_{DE}}{\rho_{m0}}= \frac{H^2}{ {\bar m}^{2}}-g(a)- \chi a^{-4} , \label{50}
\end{eqnarray}
\begin{eqnarray}
y_R \equiv  \frac{R}{ {\bar m}^{2} }- \frac{dg(a)}{d\ln a},\label{51} 
\end{eqnarray}
where $\rho_{DE}$ denotes  the energy density of dark energy, ${\bar m}^{2}$, the mass scale, given by \cite{eti13}
${\bar m}^{2} = \frac { {\kappa}^2 \rho_{m0}}{3}$.\par
Making use of (\ref{46}), the expression $\frac{1}{{\bar m}^{2}} \frac{dR} {d\ln a}$ yields
\begin{eqnarray}
\frac{1}{{\bar m}^{2}} \frac{dR} {d\ln a}= \frac{1}{H^2 f_{RR}} \bigg [ \frac{f_T}{3 {\bar m}^{2}} (\rho_ {matt}+P_{matt})+
\frac{1}{3{\bar m}^{2}}\rho_ {matt}-\frac{H^2}{{\bar m}^{2}}+\frac{1}{6{\bar m}^{2}}(f-R)-(1-f_R)\bigg(\frac{H}{{\bar m}^{2}} \frac{dH}{d\ln a}+\frac{H^2}{{\bar m}^{2}}\bigg)\bigg].\
\label{52}
\end{eqnarray}
We recall in this work that the trace of the stress tensor depends on the nature of the matter content. Therefore, using the same model (\ref{000}) where  $ T = \rho_{m0} [g(a)(1-3w)]$, and also  (\ref{0000}) and (\ref{45}), Eq.~(\ref{52}) becomes
\begin{eqnarray}
\frac{1}{{\bar m}^{2}} \frac{dR} {d\ln a}=\frac{1}{H^2 f_{RR}}\bigg[3^{\beta} {\bar{m}}^{2\beta}(1+\beta) 
{\bigg(g(a)(1-3w)\bigg)}^{\beta}
\bigg((1+w)g(a)+\frac{4}{3}\chi a^{-4}\bigg)-y_H+\frac{1}{6{\bar m}^{2}}(f-R)- \nonumber \\
(1-f_R)\bigg(\frac{H}{{\bar m}^{2}} \frac{dH}{d\ln a}+\frac{H^2}{{\bar m}^{2}}\bigg)\bigg].
\label{53}
\end{eqnarray}
where $ \beta = \alpha -1$. Eqs. (\ref{47}) and (\ref{53}) are reduced to a coupled set of ordinary differential
 \begin{eqnarray}
 \frac{dy_H}{d\ln a}= -\frac{1}{3} y_R- 4 y_H -\frac{4}{3}\frac{dg(a)}{d\ln a}-4g(a), \label{54} 
 \end{eqnarray}
\begin{eqnarray}
\frac{dy_R}{d\ln a}=-\frac{d^2 g(a)}{{d\ln a}^{2}}+ \frac{1}{f_{RR}{\bar{m}}^2(y_H+g(a)+\chi a^{-4})}
\bigg[3^{\beta} {\bar{m}}^{2\beta}(1+\beta)
{\bigg(g(a)(1-3w)\bigg)}^{\beta}\bigg((1+w)g(a)+\frac{4}{3}\chi a^{-4}\bigg)- \nonumber \\
y_H+\frac{1}{6{\bar m}^{2}}(f-R)-(1-f_R)\bigg(\frac{1}{2} \frac{dy_H}{d\ln a}+
\frac{1}{2}\frac{dg(a)}{d\ln a}+y_H+g(a)-\chi a^{-4} \bigg) \bigg] . 
\label{55}
\end{eqnarray}
 Moreover, the curvature scalar is expressed as
 \begin{eqnarray}
 R = -3{\bar m}^{2}\bigg(4y_H+4g(a)+\frac{dy_H}{d\ln a}+ \frac{dg(a)}{d\ln a}\bigg).\label{56}
 \end{eqnarray}
 By operating the differentiation of the relation (\ref{54}) with respect to $\ln a $, and  eliminating $ \frac{dy_R} {d\ln a} $ from this result and (\ref{55}), we obtain
 \begin{eqnarray}
\frac{d^2 y_H}{{d\ln a}^{2}}+\bigg(4-\frac{1-f_R}{6{\bar m}^{2}f_{RR}(y_H+g(a)+\chi a^{-4})}\bigg) \frac{dy_H}{d\ln a}+
\bigg(\frac{f_R-2}{3{\bar m}^{2}f_{RR}(y_H+g(a)+\chi a^{-4}) }\bigg) y_H+ 
\bigg[\frac{d^2 g(a)}{{d\ln a}^{2}}+4\frac{dg(a)}{d\ln a}+\nonumber  \\
\frac{3^{\beta+1} {\bar{m}}^{2\beta}2(1+\beta){\bigg(g(a)(1-3w)\bigg)}^{\beta} \bigg( (1+w)g(a)+\frac{4}{3} \chi a^{-4}\bigg)+
(f_R-1)\bigg(3\frac{dg(a)}{d\ln a}+6g(a)-6\chi a^{-4}+\frac{f-R}{{\bar{m}}^{2}}\bigg)}{ f_{RR}{\bar{m}}^{2}(y_H+g(a)+\chi a^{-4})}\bigg]=0.
\label{57}
\end{eqnarray}
Taking into account the relations
\begin{eqnarray}
\frac{d}{d\ln a}=-(1+z)\frac{d}{dz}, \label{58} 
\end{eqnarray}
\begin{eqnarray}
\frac{d^2}{{d\ln a}^{2}} = {(1+z)}^{2}\frac{d^2}{{dz}^{2}}+(1+z)\frac{d}{dz},  \label{59} 
\end{eqnarray}
we express the equation (\ref{57}) in terms of the redshift, as follows
\begin{eqnarray}
\frac{d^2 y_H}{dz^2}+J''_1 \frac{dy_H}{dz}+J''_2 y_H +J''_3  = 0, \label{60}   
\end{eqnarray}
where
\begin{eqnarray}
J''_1 = \frac{1}{(1+z)}\bigg[-3+ \frac{f_R-2}{6{\bar{m}}^{2}f_{RR}(y_H+g(z)+\chi{(1+z)}^{4})}\bigg], \label{61}  
\end{eqnarray}
\begin{eqnarray}
J''_2 = \frac{1}{(1+z)^2} \bigg[ \frac{f_R}{3{\bar{m}}^{2}f_{RR}(y_H+g(z)+\chi{(1+z)}^{4})}\bigg], \label{62} 
\end{eqnarray}
 \begin{eqnarray}
 J''_3 = \frac{d^2 g(z)}{dz^2}-\frac{3}{(1+z)}\frac{dg(z)}{dz}+ \frac{1}{{(1+z)}^{2}6{\bar{m}}^{2} f_{RR}(y_H+g(z)+
 \chi{(1+z)}^{4})}\nonumber \\ 
 \Bigg[ 3^{\beta} {\bar{m}}^{2\beta}2(1+\beta) {\bigg (g(z)(1-3w)\bigg)}^{\beta}\bigg( g(z)(1+w)+\frac{4}{3}\chi {(1+z)}^4
 \bigg)+\\ \nonumber
 f_R \bigg (-(1+z)\frac{dg(z)}{dz}+2g(z)-2\chi{(1+z)}^{4} \bigg)+\frac{f(R)}{3{\bar{m}}^{2}}+2g(z)+
 2\chi {(1+z)}^{4}+3^{\beta} {\bar{m}}^{2\beta}{\bigg(g(z)(1-3w)\bigg)}^{\beta+1} \Bigg]. 
 \label{63}
\end{eqnarray}
Eq.~(\ref{60}) characterises the equation to be used for describing the cosmological evolution of the dark energy in  the universe filled with collisional matter and radiation.
We also solve numerically this equation  for a given $g(a)$ from (\ref{49}) and we present the cosmological evolution of the dark energy scale $ y_H \equiv \frac{\rho_{DE}}{\rho_{m0}} $ of collisional matter $(w=0.6)$ in comparison to the non-collisional matter as functions of the redshift.  

\begin{figure}[h]
\centering
\begin{tabular}{rl}
\includegraphics[width=7cm, height=7cm]{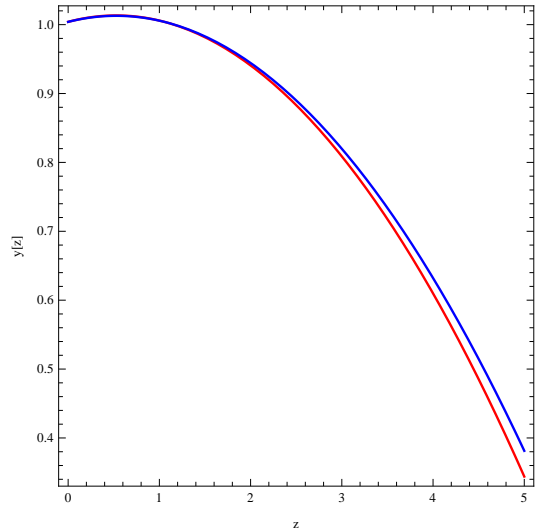}
\end{tabular}
\caption{Cosmological evolutions of $ y_H $ as functions of the redshift $z$ for $ b=0.19$, $Rs=1.5$ of
collisional matter (blue) and the non-collisional matter (red).}
\label{fig5}
\end{figure}

In this rubric, we plot the evolution of the parameter of equation of state for dark energy $w_{DE}(z) \equiv P_{DE}/ \rho_{DE}$, given by 
\begin{eqnarray}
w_{DE}(z) = -1+\frac{1}{3}(1+z)\frac{1}{y_H}\frac{d y_H}{dz},\label{64}
\end{eqnarray}
derived by the continuity equation
\begin{eqnarray}
\dot{\rho_{DE}}+3H(1+w_{DE})\rho_{DE} = 0, \label{65}
\end{eqnarray}
for collisional matter $(w=0.6)$ in comparison to the non-collisional one.\par
As performed in the previous section, we are also interested to the evolution of others parameters. Theses parameters, the Hubble parameter $H(z)$, the curvature scalar $R(z)$ and the parameter $w_{eff}$ of the effective  equation of state,  can be described by still considering that the universe is filled by collisional matter and radiation. In the same way, we find the numerical solution for  $ y_H $ (\ref{60}) for both collisional and non-collisional matters, and compare them. These parameters can be expressed as follows 
\begin{eqnarray}
H(z) = \sqrt{{\bar m}^{2}(y_H+g(z)+\chi{(1+z)}^{4})} \label{66} 
\end{eqnarray}
obtained from equation (\ref{50}),
\begin{eqnarray}
R = 3{\bar m}^{2} \bigg(4y_H + 4g(z)-(1+z)\frac{dy_H}{dz}-(1+z)\frac{dg(z)}{dz} \bigg) \label{67} 
\end{eqnarray}
obtained by combining the equations (\ref{56}) and (\ref{58}),
\begin{eqnarray}
w_{eff} = -1+\frac{2(1+z)}{3H(z)} \frac{dH(z)}{dz}.\label{68} 
\end{eqnarray}

 \begin{figure}[h]
 \centering
 \begin{tabular}{rl}
 \includegraphics[width=7cm, height=7cm]{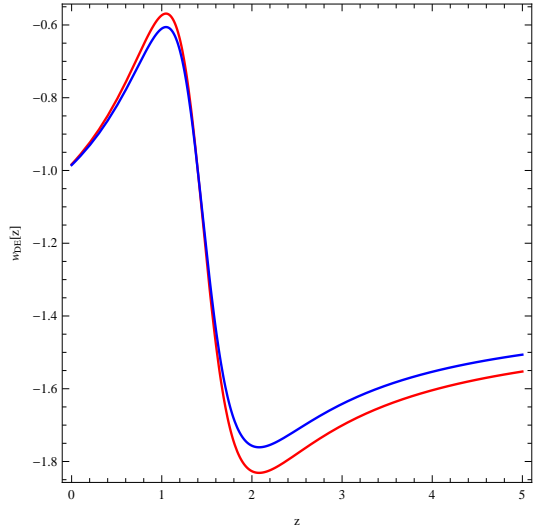}
 \end{tabular}
 \caption{Cosmological evolutions of $\omega_{DE}$ as functions of the redshift z for $ b=0.19$, $Rs=1.5$ of collisional matter for w=0.6 (blue)
 and the non-collisional matter (red).}
 \label{fig6}
 \end{figure}

\newpage
 
  \begin{figure}[h]
   \centering
   \begin{tabular}{rl}
   \includegraphics[width=7cm, height=7cm]{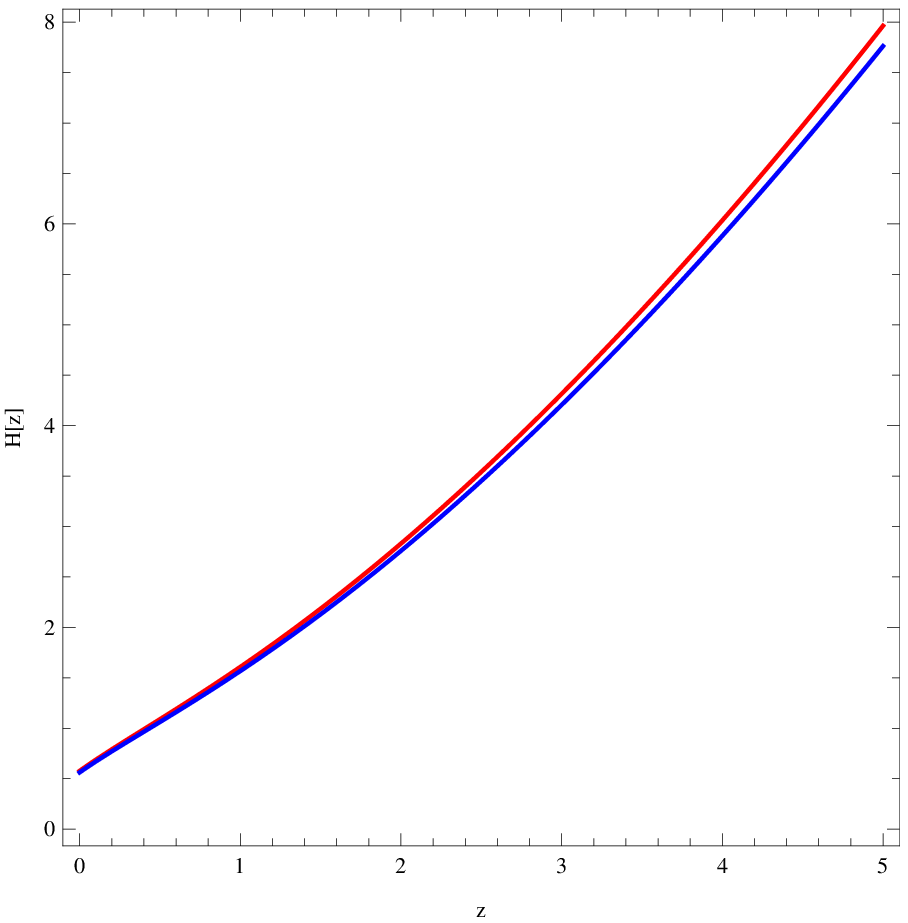}&
   \includegraphics[width=7cm, height=7cm]{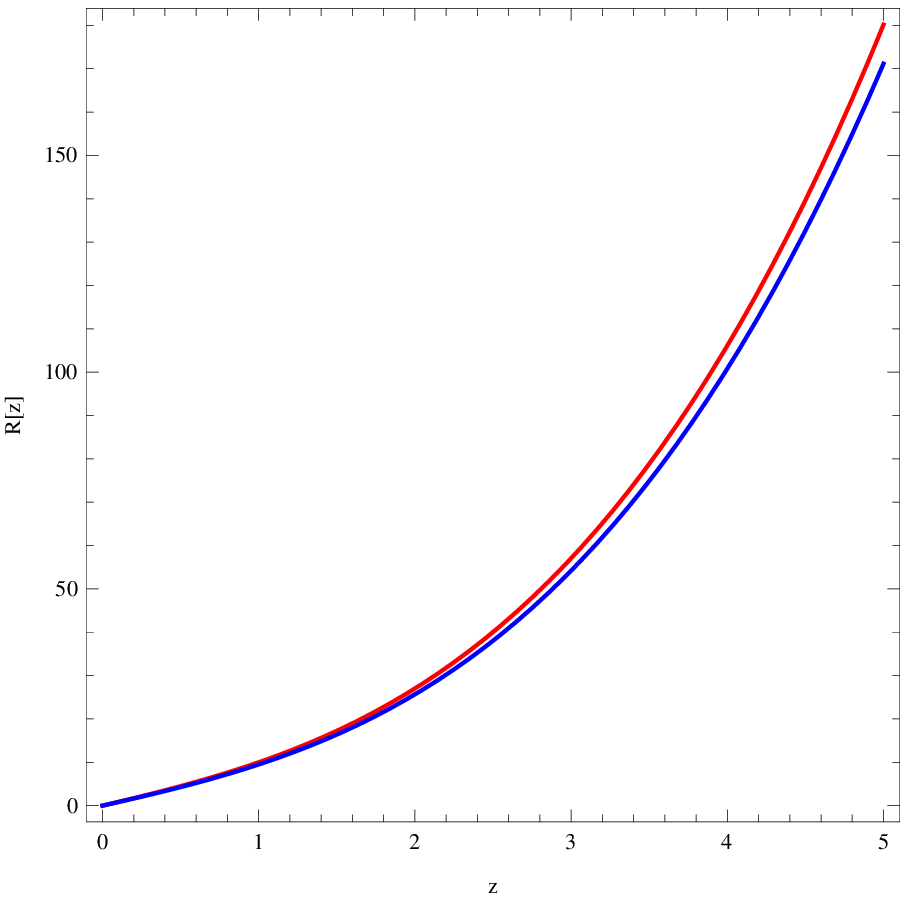}
   \end{tabular}
   \caption{Comparison of the Hubble parameter $H(z)$ over z (left) and of the Ricci scalar
  $R(z)$ (right), for $ b=0.19 $, $Rs=1.5$. The red line corresponds to non-collisional matter while the
  blue corresponds to collisional matter for w=0.6.}
  \label{fig7}
  \end{figure}

\begin{figure}[h]
\centering
\begin{tabular}{rl}
\includegraphics[width=7cm, height=7cm]{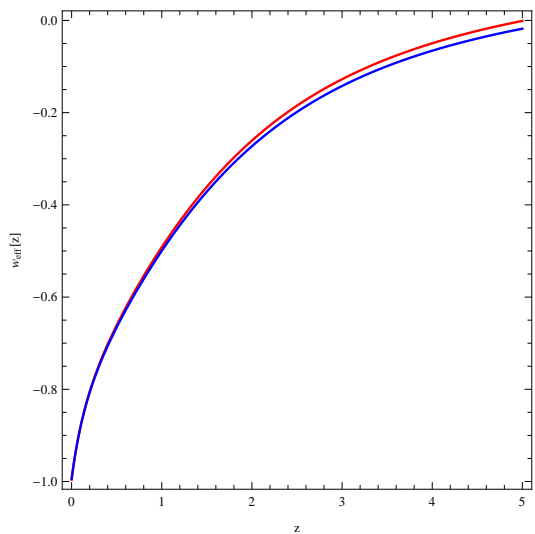}
\end{tabular}
\caption{Cosmological evolutions of $w_{eff}$ as functions of the redshift z for $ b=0.19$, $Rs=1.5$ of collisional matter for w=0.6 (blue)
and the non-collisional matter (red).}
\label{fig8}
\end{figure}

For the Figs.~$5$, $6$, $7$ and $8$, curves traducing the evolution of $y_H(z)$, $w_{DE}(z)$, $H(z)$, $R(z)$ and $w_{eff}(z)$, present some values strongly consistent with the cosmological observational data. These features prove that the fact of considering the coexistence of the collisional matter and radiation do not change the well know behaviour the different parameters, and confirm that such consideration may be made.  

\section{Conclusion} \label{sec6}

In this paper we undertake the $f(R,T)$ theory of gravity. We focus our attention on the cosmological evolution 
of the parameter of deceleration $q(z)$ and also the parameter $w_{eff}$ of the effective equation of state. 
The spacial aspect in this paper is that, besides the usual ordinary and the dark energy, new matter is assumed 
to participate to the matter contribution of the universe. This matter is self-interacting and called collisional
matter, but with positive pressure. What we now generally is that the ordinary matter interacts with the dark 
energy, but any of them does self interacts. From the cosmological considerations, its quite reasonable to 
introduce such a matter, i.e, a collisional, and search for its effects on the cosmological evolution of the universe.\par
To this end, we considered two $f(R,T)$ models as a sum of two gravitational models $f_1(R)$ and $f_2(T)$, respectively
two functions depending the curvature scalar $R$ and the trace $T$ of the energy-momentum tensor. The general form of the $f(R,T)$ models is constrained to the conservation of the energy leading to  the form (\ref{000}). The task has been to choose a suitable expression for the function $f(R)$. Therefore, we considered, first, a generalized power-law in $R$, and the second expression is an exponential form in $R$. We plot the evolutions of $q(z)$ and $w_{eff}$ for both the collisional matter, non-collisional matter and the $\Lambda$CDM and compare them. Our results show  that, depending on the type of the model under consideration, the curve characteristic of the collisional matter approaches the $\Lambda$CDM one that the curve that describes the evolution of the dust. In the same way, the curve representative of $w_{eff}$ shows that its values are consistent with the observational, confirm 
the importance of considering the collisional matter in the study of the evolution of our universe.\par 
Moreover, we point out other type of matter, the so-called cardassian self-interacting matter, possessing a negative
pressure and its corresponding parameter of equation of state has to be different from $-1$. This means that if such 
a matter is included in the content of the universe, there does not have vacuum energy. In this case, the same analysis 
have been done, as in the case of collisional matter, comparing the curves representative of this later to those of the
dust and also the $\Lambda$CDM. Our results show that, for both the first and the second $f(R,T)$ models, there is quite
reasonable to consider the existence of the cardassian matter since the curve characteristic of this latter approached more
the $\Lambda$CDM  that the dust for some values of the redshift, confirming also that this kind of matter may be considered
as a component of the universe, because also in this, the parameter $w_{eff}$ takes values strongly consistent with the 
observational data.

\vspace{1cm}

{\bf Acknowledgments}: 
The authors thank Prof. S. D. Odintsov for useful comments and suggestions. E. H. Baffou  thanks
IMSP for every kind of support during the realization of this work. Manuel E. Rodrigues  
thanks UFPA, Edital 04/2014 PROPESP, and CNPq, Edital MCTI/CNPQ/Universal 14/2014,  for partial financial support. M.J.S. Houndjo and A.V.Kpadonou thank {\it Ecole Normale Sup\'erieure de Natitingou} for partial financial support.

\appendix
\section{The energy-momentum tensor for a perfect fluid}

{\bf Let us assume that the Lagrangian density  $\mathcal{L}_m$ is a sum of the kinetic term  $\mathcal{L}_{mK}$ and the one linked to the internal energy namely $\mathcal{L}_{mU}$, i.e, 
\begin{equation}
\mathcal{L}_m = \mathcal{L}_{mK}+\mathcal{L}_{mU}.
\end{equation}
Therefore, the energy-momentum tensor $T_{\mu\nu}$, reads
\begin{equation}
T_{\mu\nu}=-\frac{2}{\sqrt{-g}}\Biggl[\frac{\delta\left(\sqrt{-g}\mathcal{L}_{mK}\right)}{\delta g^{\mu\nu}}+\frac{\delta\left(\sqrt{-g}\mathcal{L}_{mU}\right)}{\delta g^{\mu\nu}} \Biggr] \\
= T_{\mu\nu(K)}+T_{\mu\nu(U)}
\end{equation}
The well known expression of the kinetic energy-momentum tensor is expressed as $ T_{\mu\nu(K)}=-\rho u_{\mu}u_{\nu}$.\\
Let us consider $\mathcal{L}_{mU}$ as
\begin{equation}
\mathcal{L}_{mU}= f(\lambda, u_{k})= -p\sqrt{g^{\mu\nu}u_{\mu}u_{\nu}}+\lambda\biggl(\sqrt{g^{\mu\nu}u_{\mu}u_{\nu}}-1\biggr).
\end{equation}
Using the method of Lagrange multipliers, the equations for the parameter $\lambda$ and the variables $u_{k}$ are given by  
\begin{equation}
\frac{\partial f}{\partial{\lambda}} = \sqrt{g^{\mu\nu}u_{\mu}u_{\nu}}-1
\end{equation}
and
\begin{equation}
\frac{\partial f}{\partial{u_{k}}}= \frac{1}{2}(\lambda-p)\sqrt{g^{\mu\nu}u_{\mu}u_{\nu}}2g^{k\mu}u_{\mu}.
\end{equation}
By setting $\frac{\partial f}{\partial{\lambda}}=0$ and $\frac{\partial f}{\partial{u_{k}}}=0$, one gets
$ \sqrt{g^{\mu\nu}u_{\mu}u_{\nu}}=1$ and $\lambda=p$. \\
Therefore, one can rewritten $\mathcal{L}_{mU}$ as
\begin{equation}
\mathcal{L}_{mU}=-p+p\biggl( \sqrt{g^{\mu\nu}u_{\mu}u_{\nu}}-1\biggr) =-p
\end{equation}
The internal energy-momentum tensor $T_{\mu\nu(U)}$ can be developed in the following from
\begin{equation}
T_{\mu\nu(U)}=  g_{\mu\nu}\mathcal{L}_{mU}-\frac{{2}{\partial{\mathcal{L}_{mU}}}}{\partial{g^{\mu\nu}}} 
\end{equation}
The second term $\frac{{\partial{\mathcal{L}_{mU}}}}{\partial{g^{\mu\nu}}}$ can be expressed as
\begin{equation}
\frac{{\partial{\mathcal{L}_{mU}}}}{\partial{g^{\mu\nu}}} = -\frac{\partial p}{\partial{g^{\mu\nu}}}+ \biggl(\sqrt{g^{\mu\nu}u_{\mu}u_{\nu}}-1\biggr)\frac{\partial p}{\partial{g^{\mu\nu}}}+\\
p \frac{\partial}{\partial{g^{\mu\nu}}}\biggl(\sqrt{g^{\mu\nu}u_{\mu}u_{\nu}}-1\biggr) .
\end{equation}
 The pressure $p$ does not depend on the metric tensor $g^{\mu\nu}$, such that  $\frac{\partial p}{\partial{g^{\mu\nu}}}=0$,
 and then $\frac{{\partial{\mathcal{L}_{mU}}}}{\partial{g^{\mu\nu}}}$ reduces to
 \begin{equation}
 \frac{{\partial{\mathcal{L}_{mU}}}}{\partial{g^{\mu\nu}}}= \frac{1}{2}pu_\mu u_\nu 
 \end{equation}
and finally, the internal energy-momentum tensor becomes
\begin{equation}
 T_{\mu\nu(U)}= -p\biggl(u_\mu u_\nu-g_{\mu\nu} \biggr)
\end{equation}
By using the contribution of the kinetic and the internal energy-momentum tensors and absorbing the sign $(-)$ into the total stress tensor, one gets the energy-momentum tensor of matter fluid as
\begin{equation}
T_{\mu\nu}=\left(\rho+p \right)u_{\mu}u_{\nu}-p g_{\mu\nu}. 
\end{equation}
It comes from the above calculus that the expression of the energy-momentum tensor is a consequence of the fact of considering that the pressure does not depends on the metric tensor $g^{\mu\nu}$. }


\end{document}